\RequirePackage{fix-cm}
\documentclass[smallcondensed]{svjour3}     
\smartqed  
\usepackage{graphicx}
\usepackage{amsmath}
\usepackage{amsfonts}
\usepackage{graphicx} 
\usepackage{amsmath,esint} 
\usepackage{multirow}
\usepackage{color}
\usepackage{notoccite}
\usepackage{bm}
\usepackage{url} 
\usepackage{epstopdf}

\newcommand\lrcurs{{\mbox{$\resizebox{.14in}{.075in}{\includegraphics{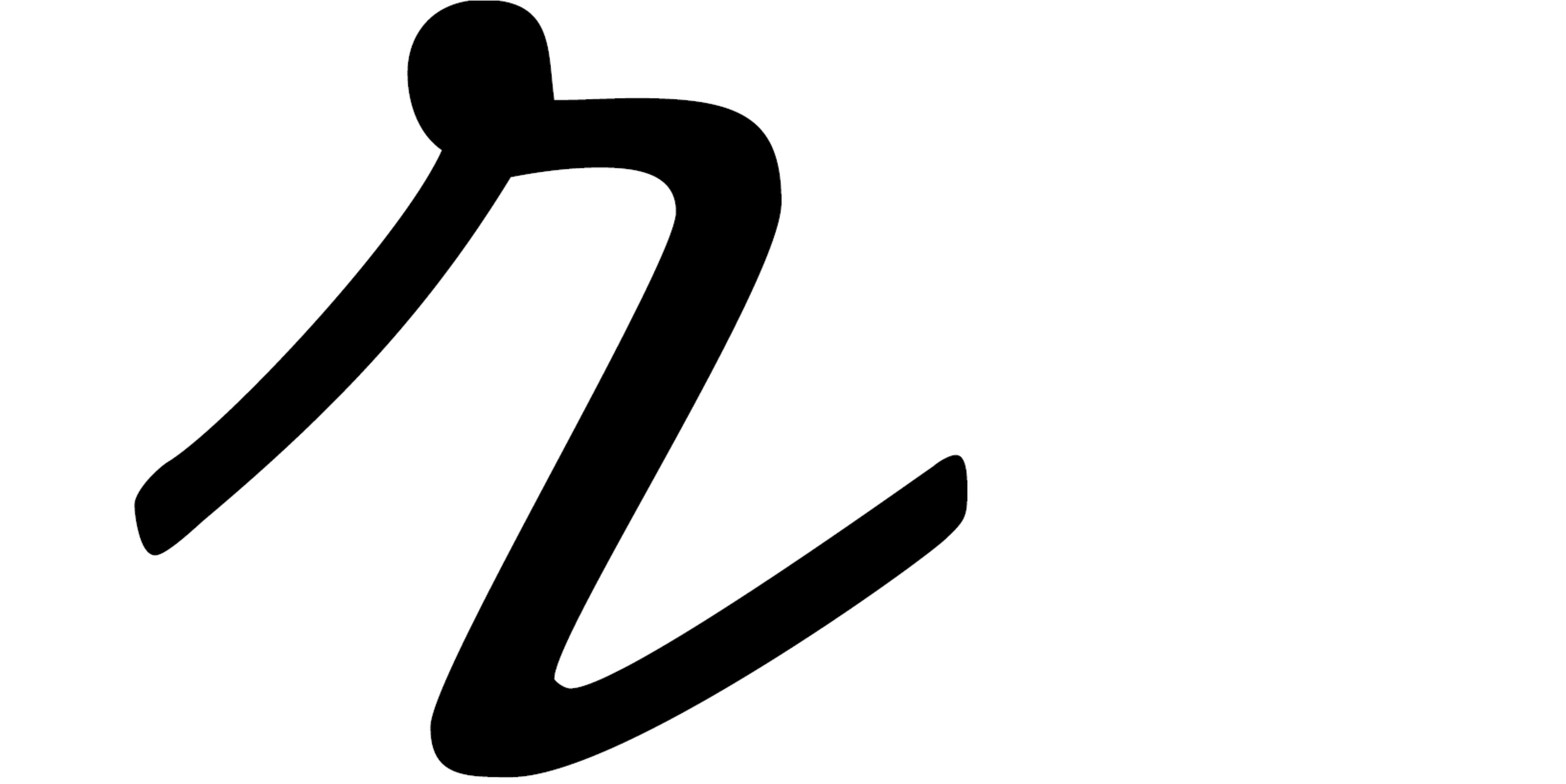}}$}}}

\newcommand\lbrcurs{{\mbox{$\resizebox{.14in}{.075in}{\includegraphics{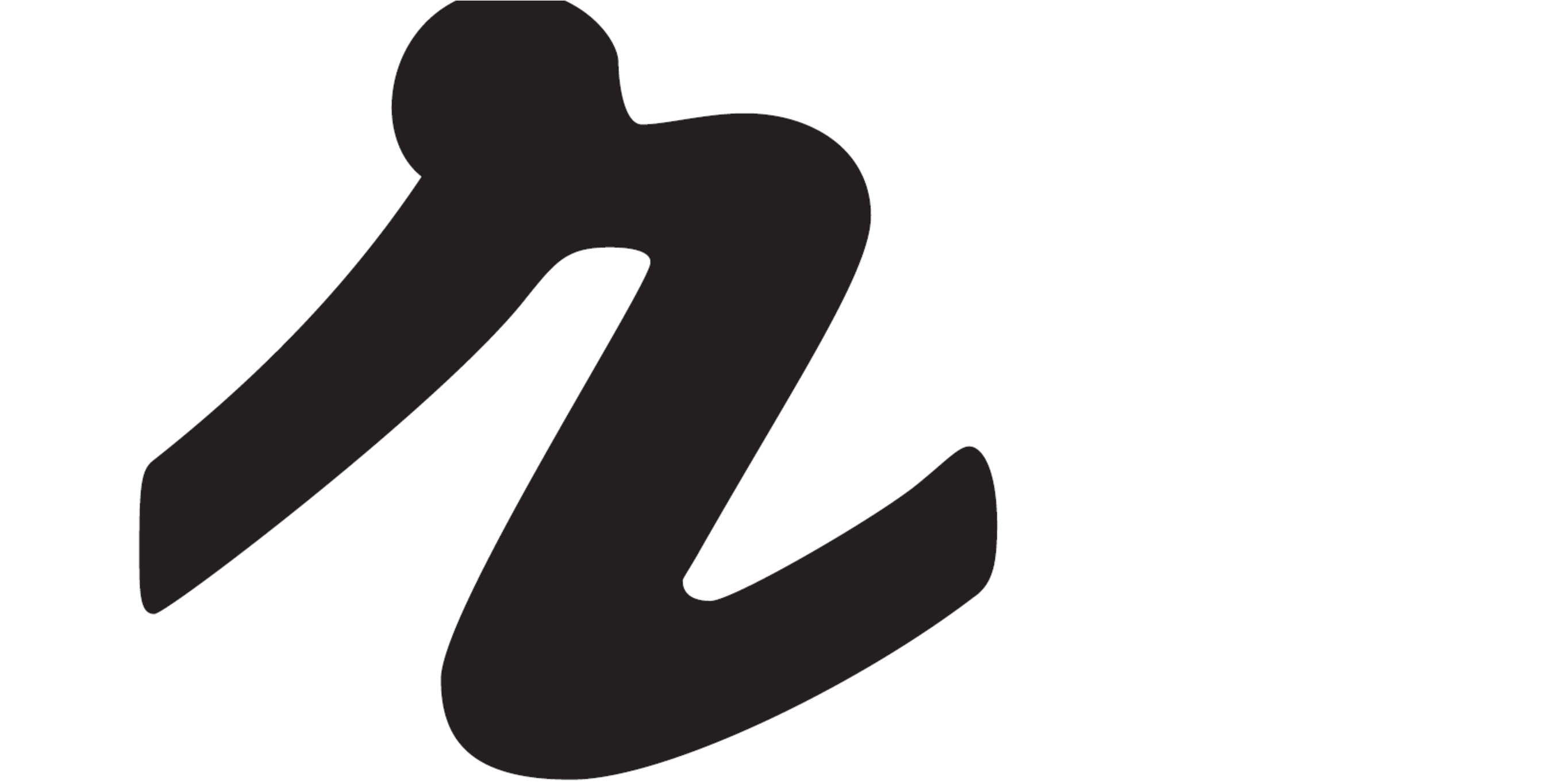}}$}}}

\begin{document}

\title{Long-Range Longitudinal Electric Wave in Vacuum Radiated by Electric Dipole: Part I 
}


\author{Altay Zhakatayev, Leila Tlebaldiyeva}


\institute{Altay Zhakatayev \at
	School of Science and Technology, \\
	Leila Tlebaldiyeva \at
	School of Engineering, Nazarbayev University \\
	Room 402, Block 7, Kabanbay Batyr Ave 53 \\
	Astana, Kazakhstan, Z05H0P9 \\
              Tel.: +77172 709097\\
              \email{azhakatayev@nu.edu.kz}           
           }
\date{Received: date / Accepted: date}

\maketitle

\begin{abstract}
 
In this work by using the assumptions that wavelength is much smaller than charge separation distance of an electric dipole, which in turn is much smaller than a distance up to the point of observation, the new results for radiation of an electric dipole were obtained. These results generalize and extend the standard classical solution, and they indicate that under the above assumptions the electric dipole emits both long-range longitudinal electric and transverse electromagnetic waves. For a specific values of the dipole system parameters the longitudinal and transverse electric fields are displayed. Total power emitted by electric and electromagnetic waves are calculated and compared. It was shown that under the standard assumption of charge separation distance being much smaller than wavelength: a) classical solution correctly describes the transverse electromagnetic waves only; b)  longitudinal electric waves are non-negligible; c) total radiated power is proportional to the fourth degree of frequency and to the second degree of the charge separation distance; d) transverse component of our solution reduces to classical solution. In case wavelength is much smaller than charge separation distance: a) the classical solution is not valid and it overestimates the total radiated power; b) longitudinal electric waves are dominant and transverse electromagnetic waves are negligible; c) total radiated power is proportional to the third degree of frequency and to the charge separation distance; d) most of the power is emitted in a narrow beam along the dipole axis, thus emission of waves is focused as with lasers.
 
\end{abstract}

\section{Introduction}

Radiation of the long-range electromagnetic waves, emitted by a classical electric dipole, is well known. Derivation can be found in almost any textbook on electrodynamics \cite{JerroldFranklin,WalterGreiner,DavidGriffiths,Ohanian1988,WolfgangPanofsky}. The main assumptions used in the derivation of these results are
\begin{equation}\label{eqn:old_assumption}
d \ll \lambda \ll r
\end{equation}
where $d$ is the distance between two point charges of the electric dipole, $\lambda$ is wavelength of the emitted electrodynamic wave and $r$ is the distance from midpoint of the dipole up to the observation point.
We will consider radiation by the electric dipole, but with the following assumptions
\begin{equation}\label{eqn:new_assumption}
\lambda \ll d \ll r
\end{equation}
It turns out that these new assumptions lead to completely novel results. There are two electric dipole systems considered in the literature: one in which charges have constant magnitude but oscillating position \cite{WalterGreiner,schwinger1998classical}, and another in which charges are fixed in space and so do not move, but their charge magnitudes are oscillating \cite{DavidGriffiths,heald1995classical}. In this paper our focus will be on the latter system, and the spherical coordinate system is utilized, where radial field components will be called longitudinal. To the best knowledge of authors, these results were not published before. 

Literature covering the longitudinal electric waves is scarce. This is due to the fact that according to Gauss's law longitudinal electric waves are impossible in vacuum (for the spherical waves). However in this paper we challenge the implications of the Gauss's law with respect to the existence of longitudinal electric waves by considering the electric dipole radiation. A theoretical method for quantizing the radiation field using transverse, longitudinal and scalar photons was introduced in \cite{Gupta1950theory}. An attempt to consider the existence of longitudinal electric waves using general framework of classical electrodynamics was tried in \cite{vanVlaenderen2001fn}. Possible existence and physical relevance of longitudinal electromagnetic waves from quantum electrodynamic point of view was studied in \cite{Khvorostenko1992}. Formation of transient longitudinal electromagnetic wave at the instant when a point charge crosses the perfectly conducting half-space and annihilates with its image was theoretically shown in \cite{prijmenko2011formation}. 

The main contributions of the paper are the following. Firstly, radiation of the infinitesimal electric dipole with oscillating charge magnitudes, but fixed charge positions, was generalized for $\frac{d}{\lambda} \in \mathbb{N}$. Secondly, radiation due to electric current wave traveling in one direction was obtained. In essence, traveling wave dipole (antenna) is considered. Most of the linear antenna theories consider standing wave currents, which are composed of two current waves traveling in the opposite directions. Thirdly, contribution to radiation of the linear charge densities, generated along the thin wire due to sinusoidal current, was considered. Finally, power generated by longitudinal waves was calculated and compared with the transverse power.

\section{Theoretical Derivation}\label{sec:theoretical_methodology}

\begin{figure}
	\centering
	\includegraphics[scale = 0.8]{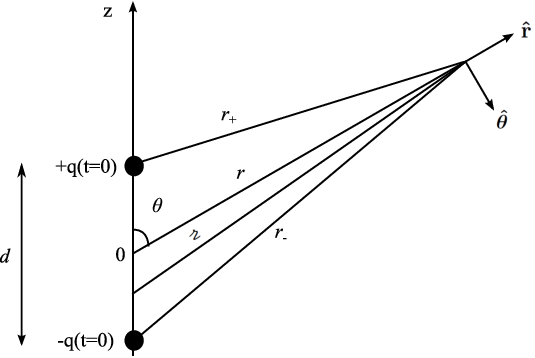}
	\caption{Schematic diagram of an electric dipole}
	\label{fig:dipole}
\end{figure}

Scalar and vector potentials are defined using charge $\rho$ and current $\mathbf{J}$ densities as
\begin{subequations}
	\begin{equation}
	V(\mathbf{r},t) = \frac{1}{4\pi\epsilon_0} \int \frac{\rho(\mathbf{r}',t_r)}{\lrcurs} dV'	
	\end{equation}
	\begin{equation}
	\mathbf{A}(\mathbf{r},t) = \frac{\mu_0}{4\pi}\int \frac{\mathbf{J}(\mathbf{r}',t_r)}{\lrcurs} dV'
	\end{equation}
\end{subequations}
here $\lbrcurs$ is radius vector from charge location at retarded time $t_r$ up to the point, where potential fields are being observed, and $\lrcurs$ is its magnitude. $t_r$ is ``retarded'' time, i.e. time at which scalar potential was ``emitted'' by a charge element or vector potential was ``emitted'' by a current element. $\mathbf{r}$, $r$ and $\bf{\hat{r}}$ are correspondingly radius vector from the origin of reference coordinate system up to the point of observation, its magnitude and its unit vector. Radius vector from origin of reference system up to the point where charge or current element is located at retarded time $t_r$ is denoted as $\mathbf{r}'$ and $\mathbf{r}' = \mathbf{r}-\lbrcurs$. The point of observation is where we want to find the electric and magnetic fields. In this paper we will use spherical coordinate system, as it is used for derivation of dipole radiation in classical case, Fig. \ref{fig:dipole}. The unit vectors of the spherical frame are $\mathbf{\hat{r}}$, $\bm{\hat{\theta}}$ and $\bm{\hat{\phi}}$.

All of the derivation will be done for point charges. However in reality we should employ in the analysis small finite conductive spheres instead of point charges, due to the fact that by definition charge value of the point charges (e.g. an electron) is constant. If we assume that electric potential on the outside surface of the small conducting sphere changes simultaneously, then the analysis will not change if point charges are used instead of finite but small spheres. Therefore bear in mind that even though point charge is used in the analysis, in reality we mean finite but small spheres, charge magnitude of which can change. In other words, it might be said that in this analysis radius of small spheres is taken as negligible compared to any other spatial dimensions of the system. Similarly, cross section of the current carrying wire, which connects the two conducting spheres, is assumed to be negligible. The scalar and vector potentials of a point charge $q$ and of a steady current $\mathbf{I}$ for classical low velocity case ($\frac{v}{c} \ll 1$) are given as 
\begin{subequations}
	\begin{equation}\label{eqn:potentials1}
	V(\mathbf{r},t) = \frac{q(\mathbf{r}',t_r)}{4\pi\epsilon_0 \lrcurs}
	\end{equation}
	\begin{equation}\label{eqn:potentials2}
	\mathbf{A}(\mathbf{r},t) = \frac{\mu_0 \mathbf{I(\mathbf{r}',t_r)}}{4\pi \lrcurs}
	\end{equation}	
\end{subequations}
One point charge, which will be called ``upper'' charge $q_+$, is located at  $z=\frac{d}{2}$. The other charge, it will be given a name of ``lower'' charge $q_-$, is located at $z=-\frac{d}{2}$. Two charges are separated by a distance $d$. $r_+$ is distance from the upper charge to the point of observation, while $r_-$ is the same distance but from the lower charge, Fig. \ref{fig:dipole}. We assume that charges are stationary, i.e. not moving, but charge magnitude is a function of time. The assumptions in (\ref{eqn:new_assumption}) can be cast in the following dimensionless form
\begin{subequations}
	\begin{equation}\label{eqn:as1}
	\frac{d}{r} \ll 1
	\end{equation}
	\begin{equation}\label{eqn:as2}
	\frac{c}{\omega d} \ll 1
	\end{equation}
	\begin{equation}\label{eqn:as3}
	\frac{c}{\omega r} \ll 1
	\end{equation}
\end{subequations}
where we used the identity $\lambda = \frac{2\pi c}{\omega}$, $\omega$ is the frequency of the wave. We note that in the standard classical derivation of the electric dipole radiation inverse of assumption (\ref{eqn:as2}) is used
\begin{equation}\label{eqn:as_old}
\frac{c}{\omega d} \gg 1
\end{equation}
We can use the following identity for $r_+$ and $r_-$:
\begin{equation}
r_{\pm} = r\sqrt{1 \mp \frac{d}{r} \cos{\theta} +(\frac{d}{2r})^2}
\end{equation}
If Taylor series expansion is applied for RHS of the above equation and (\ref{eqn:as1}) is used to take only the first order terms, then we obtain the following approximations
\begin{subequations}
	\begin{equation}\label{eqn:app1}
	r_{\pm} \approx r(1\mp \frac{d}{2r}\cos{\theta})
	\end{equation}
	\begin{equation}\label{eqn:app2}
	\frac{1}{r_{\pm}} \approx \frac{1}{r}
	\end{equation}
\end{subequations}
We take the first order term in (\ref{eqn:app1}), because $r_{\pm}$ is present in the phase part (numerator) of the potential. While in (\ref{eqn:app2}) we do not take into account any of the $\frac{d}{r}$ terms, due to the long-range part (denominator) of the potential, which is needed for us. Additionally, the following approximation, obtained by using (\ref{eqn:app1}), will be useful for us later
\begin{equation}\label{eqn:app4}
\cos({\omega(t-\frac{r_{\pm}}{c})}) \approx \cos({\omega(t-\frac{r}{c})})\cos{(\frac{\omega d}{2c}\cos{\theta})} \mp \sin({\omega(t-\frac{r}{c})})\sin{(\frac{\omega d}{2c}\cos{\theta})}
\end{equation}
Let us assume that at the initial time $t_r=0$ the upper charge has positive charge $q_0$, while the lower charge has negative charge $-q_0$. Similar to~\cite{DavidGriffiths} (Chapter 11), the value of these charges changes according to cosine function for later times $t_r>0$ as 
\begin{equation}\label{eqn:charge_time}
q_{\pm}(\mathbf{r}',t_r) = \pm q_0\cos{\omega t_r}
\end{equation}
Let us find the scalar potentials due to the upper charge ( $V_+(\mathbf{r},t)$) and due to the lower charge ($V_-(\mathbf{r},t)$) at point $\mathbf{r}$ and at time $t$. According to (\ref{eqn:potentials1}) we will have
\begin{equation}\label{eqn:potential}
\begin{aligned}
& V_+(\mathbf{r},t) = \frac{q_+(\mathbf{r}',t_r)}{4\pi\epsilon_0 \lrcurs} = \frac{q_0 \cos{\omega t_r}}{4\pi\epsilon_0 \lrcurs} = \frac{q_0 \cos{\omega (t - \frac{r_+}{c})}}{4\pi\epsilon_0 r_+} \\
& V_-(\mathbf{r},t) = \frac{q_-(\mathbf{r}',t_r)}{4\pi\epsilon_0 \lrcurs} = -\frac{q_0 \cos{\omega t_r}}{4\pi\epsilon_0 \lrcurs} = -\frac{q_0 \cos{\omega (t - \frac{r_-}{c})}}{4\pi\epsilon_0 r_-}
\end{aligned}
\end{equation}
Then scalar potential field of the dipole will have the following form
\begin{equation}\label{eqn:scalar_potential}
V(\mathbf{r},t) = V_+(\mathbf{r},t) + V_-(\mathbf{r},t) = \frac{q_0}{4\pi \epsilon_0}\Big(\frac{\cos({\omega(t-\frac{r_{+}}{c})})}{r_+} - \frac{\cos({\omega(t-\frac{r_{-}}{c})})}{r_-}\Big)
\end{equation}
If (\ref{eqn:app2}) and (\ref{eqn:app4}) are used to open the brackets in the above equation, then we get
\begin{equation}\label{eqn:scalar_potential2}
V(\mathbf{r},t) \approx -\frac{2q_0}{4\pi\epsilon_0 r}\sin(\omega(t-\frac{r}{c}))\sin(\frac{\omega d}{2c}\cos{\theta})
\end{equation}
Gradient of the scalar potential in the spherical frame is
\begin{equation}\label{eqn:grad_v}
\nabla V = \frac{\partial{V}}{\partial{r}}\mathbf{\hat{r}} + \frac{1}{r}\frac{\partial{V}}{\partial{\theta}} \mathbf{\hat{\theta}}
\end{equation}
The scalar potential (\ref{eqn:scalar_potential2}) does not depend on $\phi$ or $\bm{\hat{\phi}}$, and therefore its gradient contains only two terms. Let us evaluate the first term in the gradient by differentiating (\ref{eqn:scalar_potential2}) and then applying (\ref{eqn:as3})
\begin{equation}
\frac{\partial{V}}{\partial{r}} \approx \frac{2q_0}{4\pi\epsilon_0 r} \frac{\omega}{c}\sin(\frac{\omega d}{2c}\cos{\theta})\cos(\omega(t-\frac{r}{c}))
\end{equation}
The second term in the gradient is obtained again by differentiating (\ref{eqn:scalar_potential2})
\begin{equation}
\frac{1}{r}\frac{\partial{V}}{\partial{\theta}} = \frac{2q_0}{4\pi\epsilon_0 r^2} \frac{\omega d}{2c}\sin{\theta}\cos(\frac{\omega d}{2c}\cos{\theta})\sin(\omega(t-\frac{r}{c}))
\end{equation}
Now we can combine above two identities, apply (\ref{eqn:as1}) to the resultant expression, and finally we get gradient of the scalar potential
\begin{equation}\label{eqn:grad_v2}
\nabla V(\mathbf{r},t) = \frac{2q_0}{4\pi\epsilon_0 r} \frac{\omega}{c} \sin(\frac{\omega d}{2c}\cos{\theta})\cos(\omega(t-\frac{r}{c})) \mathbf{\hat{r}}
\end{equation}
Current can be defined as  
\begin{equation}\label{eqn:current_time_old}
\mathbf{I}(\mathbf{r}',t_r) = \dot{q}(t_r)\mathbf{\hat{z}} = -q_0 \omega \sin{\omega t_r}\mathbf{\hat{z}}
\end{equation}
In case of classical dipole radiation, magnitude of the current $|\mathbf{I}| = q_0 \omega$ is taken as constant over distance $d$ due to assumption ($\frac{c}{\omega d} \gg 1$) \cite{Jefimenko1989,Jefimenko2000} and the only variable in the current is the retarded time $t_r$.
Above assumption intuitively can be understood as the following. The assumption $\lambda \gg d$ means that distance between charges is much smaller than wavelength of the AC current. If we ``freeze'' AC current wave in time, then different parts of the wavelength of the current wave correspond to different current magnitudes. Therefore at a given fixed time current magnitude between two charges is almost constant in space. Thus it is reasonable to assume that at any given time current magnitude between two charges of the dipole is constant. 

In this case, however, we can no longer assume that current magnitude is constant over $d$ due to (\ref{eqn:as2}). In other words, current will vary due to variation of its magnitude in space and due to variation of retarded time. Therefore we have to consider the following current 
\begin{equation}\label{eqn:current_time}
\mathbf{I}(\mathbf{r}', t_r) = -q_0 \omega \sin{(\frac{2\pi}{\lambda}(\frac{d}{2}-z)+\omega t_r)} \mathbf{\hat{z}}
\end{equation}
which simplifies to old classical expression (\ref{eqn:current_time_old}) under assumption $\lambda \gg d$ taken for classical charge. This current, similar to its classical counterpart (\ref{eqn:current_time_old}), satisfies the charge conservation law for the upper charge. In effect, (\ref{eqn:current_time}) represents AC current wave in space and in time, ``flowing'' or traveling in the direction of positive charge. As a result, here we consider traveling wave antenna (dipole). We decided to consider sine wave current because according to~\cite{balanis2005antenna} [page 163]: ``sine wave currents are more accurate representations of the current distribution of a wire antenna with arbitrary length''. Note also that we assume that electric current, or electricity, is traveling along the wire connecting the point charges at the speed $c$. This means that velocity factor of the wire connecting the point charges is taken to be 100\%. We note that the non-constant (time-varying) currents, similar to the one in (\ref{eqn:current_time}), are considered in almost any textbook about linear antenna theories or transmission line theories. For example, dipole antenna with standing wave current consisting of four traveling constant current waves was considered in \cite{zangwill2013modern}, triangular and sine standing wave currents in the small and finite dipole antenna were considered in \cite{balanis2005antenna}, sine wave current in linear antenna was considered in \cite{schwinger1998classical}. 

Next the vector potential needs to be evaluated
\begin{equation}\label{eqn:vector_potential}
\mathbf{A}(\mathbf{r},t) = \frac{\mu_0}{4\pi} \int_{-\frac{d}{2}}^{\frac{d}{2}} \frac{\mathbf{I}(\mathbf{r}',t_r)}{\lrcurs} dz = -\frac{\mu_0 q_0 \omega}{4\pi} \mathbf{\hat{z}} \int_{-\frac{d}{2}}^{\frac{d}{2}} \frac{\sin(\frac{2\pi}{\lambda}(\frac{d}{2}-z)+\omega(t-\frac{\lrcurs}{c}))}{\lrcurs} dz
\end{equation}
here $\lrcurs$ is distance from current segment, which is being integrated, up to the point of observation of fields (see Fig. \ref{fig:dipole}), $z$ is integration variable, $-\frac{d}{2} \leq z \leq \frac{d}{2}$. For current segment we can write similar approximations as for the two point charge locations
\begin{subequations}
	\begin{equation}\label{eqn:app5}
	\lrcurs \approx r(1\mp \frac{z}{r}\cos{\theta})
	\end{equation}
	\begin{equation}\label{eqn:app6}
	\frac{1}{\lrcurs} \approx \frac{1}{r}
	\end{equation}
\end{subequations}
here the sign depend on whether $z$ is above $0$ or below it. Using trigonometric identities, sine term inside the integral (\ref{eqn:vector_potential}) can be written as a sum of four terms
\begin{equation}\label{eqn:sine}
\begin{aligned}
& \int_{-\frac{d}{2}}^{\frac{d}{2}} \sin(\frac{2\pi}{\lambda}(\frac{d}{2}-z)+\omega(t-\frac{\lrcurs}{c})) \frac{1}{\lrcurs} dz=\\ 
& \int_{-\frac{d}{2}}^{\frac{d}{2}} \Big( \sin(\frac{2\pi}{\lambda}(\frac{d}{2}-z)) \cos(\omega(t-\frac{\lrcurs}{c}))+\cos(\frac{2\pi}{\lambda}(\frac{d}{2}-z)) \sin(\omega(t-\frac{\lrcurs}{c})) \Big) \frac{1}{\lrcurs} dz = \\
& \int_{-\frac{d}{2}}^{\frac{d}{2}} \Bigg( \Big(\sin(\frac{\pi d}{\lambda}) \cos(\frac{2 \pi z}{\lambda}) - \cos(\frac{\pi d}{\lambda}) \sin(\frac{2 \pi z}{\lambda})\Big) \cos(\omega(t-\frac{\lrcurs}{c})) + \\
& \Big(\cos(\frac{\pi d}{\lambda}) \cos(\frac{2 \pi z}{\lambda}) + \sin(\frac{\pi d}{\lambda}) \sin(\frac{2 \pi z}{\lambda})\Big) \sin(\omega(t-\frac{\lrcurs}{c})) \Bigg) \frac{1}{\lrcurs} dz
\end{aligned}
\end{equation}
Then evaluating the integral in (\ref{eqn:vector_potential}) can be done by integrating separately each of the four terms in (\ref{eqn:sine}). Let us perform integration for the first term (by using (\ref{eqn:app5}) and (\ref{eqn:app6}))
\begin{equation}\label{eqn:integral1}
\begin{aligned}
& \int_{-\frac{d}{2}}^{\frac{d}{2}} \cos(\frac{2 \pi z}{\lambda}) \cos(\omega(t-\frac{\lrcurs}{c})) \frac{1}{\lrcurs} dz = \\
& \int_{-\frac{d}{2}}^{0} \cos(\frac{2 \pi z}{\lambda}) \cos(\omega(t-\frac{r}{c}(1 + \frac{z}{r}\cos{\theta}))) \frac{1}{r} dz + \\
& \int_{0}^{\frac{d}{2}} \cos(\frac{2 \pi z}{\lambda}) \cos(\omega(t-\frac{r}{c}(1 - \frac{z}{r}\cos{\theta}))) \frac{1}{r} dz  = \\
& \frac{1}{r} \cos(\omega(t-\frac{r}{c}))\Big( \int_{-\frac{d}{2}}^{\frac{d}{2}} \cos(\frac{2\pi z}{\lambda})\cos(\frac{\omega z}{c}\cos{\theta}) dz \Big) + \\
& \frac{1}{r} \sin(\omega(t-\frac{r}{c}))\Big(  \int_{-\frac{d}{2}}^{0} \cos(\frac{2\pi z}{\lambda})\sin(\frac{\omega z}{c}\cos{\theta}) dz - \int_{0}^{\frac{d}{2}} \cos(\frac{2\pi z}{\lambda})\sin(\frac{\omega z}{c}\cos{\theta}) dz \Big)
\end{aligned}
\end{equation}
Here we see that the above integral (integral of the first term in (\ref{eqn:sine})) is split into another three integral terms. 
Let us evaluate, using simple integral trigonometric identities when necessary, each of the three integral terms in (\ref{eqn:integral1}): 
\begin{equation} \label{eqn:integ1}
\begin{aligned}
& \int_{-\frac{d}{2}}^{\frac{d}{2}} \cos(\frac{2\pi z}{\lambda})\cos(\frac{\omega z}{c}\cos{\theta}) dz = \\
& \frac{\lambda}{2\pi}\Big( \frac{1}{1-\cos{\theta}} \sin(\frac{\pi d}{\lambda}(1-\cos{\theta})) + \frac{1}{1+\cos{\theta}} \sin(\frac{\pi d}{\lambda}(1+\cos{\theta})) \Big)
\end{aligned}
\end{equation}
\begin{equation} \label{eqn:integ4}
\begin{aligned}
& \int_{-\frac{d}{2}}^{0} \cos(\frac{2\pi z}{\lambda})\sin(\frac{\omega z}{c}\cos{\theta}) dz = \\
& \frac{\lambda}{4\pi}\Big( -\frac{1}{1+\cos{\theta}} (1-\cos(\frac{\pi d}{\lambda}(1+\cos{\theta}))) + \frac{1}{1-\cos{\theta}} (1-\cos(\frac{\pi d}{\lambda}(1-\cos{\theta}))) \Big)
\end{aligned}
\end{equation}
\begin{equation} \label{eqn:integ5}
\begin{aligned}
& \int_{0}^{\frac{d}{2}} \cos(\frac{2\pi z}{\lambda})\sin(\frac{\omega z}{c}\cos{\theta}) dz = \\
& \frac{\lambda}{4\pi}\Big( -\frac{1}{1+\cos{\theta}} (\cos(\frac{\pi d}{\lambda}(1+\cos{\theta}))-1) + \frac{1}{1-\cos{\theta}} (\cos(\frac{\pi d}{\lambda}(1-\cos{\theta}))-1) \Big)
\end{aligned}
\end{equation}
By substituting (\ref{eqn:integ1}) - (\ref{eqn:integ5}) into (\ref{eqn:integral1}), a new expression is obtained
\begin{equation}\label{eqn:integral1_result}
\begin{aligned}
& \int_{-\frac{d}{2}}^{\frac{d}{2}} \cos(\frac{2 \pi z}{\lambda}) \cos(\omega(t-\frac{\lrcurs}{c})) \frac{1}{\lrcurs} dz = \\
& \frac{\lambda}{2\pi r} \cos(\omega(t-\frac{r}{c})) \Bigg( \frac{1}{1-\cos{\theta}} \sin(\frac{\pi d}{\lambda}(1-\cos{\theta})) + \frac{1}{1+\cos{\theta}} \sin(\frac{\pi d}{\lambda}(1+\cos{\theta})) \Bigg) + \\
& \frac{\lambda}{2\pi r} \sin(\omega(t-\frac{r}{c})) \Bigg( \frac{1}{1-\cos{\theta}} (1-\cos(\frac{\pi d}{\lambda}(1-\cos{\theta}))) - \frac{1}{1+\cos{\theta}} (1-\cos(\frac{\pi d}{\lambda}(1+\cos{\theta}))) \Bigg)
\end{aligned}
\end{equation}
We need to remember that the above expression represents the integral of only one term out of four in (\ref{eqn:sine}). In order to save space, the detailed evaluation of the other terms is given in the appendix (Sec. \ref{sec:appendix}). The integral of the second, third and fourth terms in (\ref{eqn:sine}) are given in (\ref{eqn:appendix1_result}), (\ref{eqn:appendix2_result}) and (\ref{eqn:appendix3_result}) correspondingly. 
Let us substitute these results into (\ref{eqn:sine}), then for (\ref{eqn:vector_potential}) we get a long expression. The following simple identities will be useful for us
\begin{subequations}
	\begin{equation}\label{eqn:simple_rel1}
	\frac{\pi d}{2\lambda} = \frac{\omega d}{4 c}
	\end{equation}
	\begin{equation}\label{eqn:simple_rel2}
	\frac{\lambda}{2 \pi r} = \frac{c}{\omega r}
	\end{equation}
\end{subequations}
If the terms of the expression for $\mathbf{A}(\mathbf{r},t)$ are then grouped and simplified using trigonometric identities, relations (\ref{eqn:simple_rel1}) and (\ref{eqn:simple_rel2}) are applied, and later assumptions (\ref{eqn:as1}) and (\ref{eqn:as3}) are implemented, we obtain the result which retains the most significant terms
\begin{equation}\label{eqn:vector_potential2}
\begin{aligned}
& \mathbf{A}(\mathbf{r},t) \approx -\frac{2q_0}{4\pi \epsilon_0 r c} \mathbf{\hat{z}} \Big(\cos(\omega(t-\frac{r}{c})) (\frac{1}{1+\cos{\theta}} \sin(\frac{\omega d}{4c}(3+\cos{\theta})) \sin(\frac{\omega d}{4c}(1+\cos{\theta}))+\\
& \frac{1}{1-\cos{\theta}} \sin(\frac{\omega d}{4c}(1+\cos{\theta})) \sin(\frac{\omega d}{4c}(1-\cos{\theta}))) + \\
& \sin(\omega(t-\frac{r}{c})) (\frac{1}{1+\cos{\theta}} \cos(\frac{\omega d}{4c}(3+\cos{\theta})) \sin(\frac{\omega d}{4c}(1+\cos{\theta}))+\\
& \frac{1}{1-\cos{\theta}} \cos(\frac{\omega d}{4c}(1+\cos{\theta})) \sin(\frac{\omega d}{4c}(1-\cos{\theta}))) \Big)
\end{aligned}
\end{equation}
Equation for vector potential (\ref{eqn:vector_potential2}) is similar to equation for scalar potential (\ref{eqn:scalar_potential2}). Now we are at the position to find the time derivative of the vector potential
\begin{equation}\label{eqn:time_A}
\begin{aligned}
& \frac{\partial{\mathbf{A}}}{\partial{t}}(\mathbf{r},t) = -\frac{2q_0 \omega}{4\pi \epsilon_0 r c} (\cos{\theta}\mathbf{\hat{r}}-\sin{\theta}\bm{\hat{\theta}}) \\ & \Big(\cos(\omega(t-\frac{r}{c})) (\frac{1}{1+\cos{\theta}} \cos(\frac{\omega d}{4c}(3+\cos{\theta})) \sin(\frac{\omega d}{4c}(1+\cos{\theta}))+\\
& \frac{1}{1-\cos{\theta}} \cos(\frac{\omega d}{4c}(1+\cos{\theta})) \sin(\frac{\omega d}{4c}(1-\cos{\theta}))) - \\
& \sin(\omega(t-\frac{r}{c})) (\frac{1}{1+\cos{\theta}} \sin(\frac{\omega d}{4c}(3+\cos{\theta})) \sin(\frac{\omega d}{4c}(1+\cos{\theta}))+\\
& \frac{1}{1-\cos{\theta}} \sin(\frac{\omega d}{4c}(1+\cos{\theta})) \sin(\frac{\omega d}{4c}(1-\cos{\theta}))) \Big)
\end{aligned}
\end{equation}
where we used identity $\mathbf{\hat{z}} = \cos{\theta}\mathbf{\hat{r}}-\sin{\theta}\bm{\hat{\theta}}$.

A careful reader might notice that a current wave in (\ref{eqn:current_time}) leads to violation of the charge conservation law along the length of the linear antenna. The assumed current gives a rise to one-dimensional distributed charges along the length of the antenna. In order to preserve the charge conservation law, this distributed charge should be considered. By applying the charge conservation law $ \nabla \cdot \mathbf{J} = -\partial{\rho}/\partial{t}$ to the antenna, the following identity is found
\begin{equation}
\frac{\partial{I(\mathbf{r}',t_r)}}{\partial{z}} = -\frac{\partial{\xi(\mathbf{r}',t_r)}}{\partial{t_r}}
\end{equation}
where $I = \mathbf{I}(\mathbf{r}', t_r) \cdot \mathbf{\hat{z}}$ ($\mathbf{I}(\mathbf{r}', t_r)$ is given in (\ref{eqn:current_time})), while $\xi$ is the one-dimensional charge distribution along the antenna length, which can be found as
\begin{equation}\label{eqn:charge_dist}
\xi(\mathbf{r}',t_r) = -\frac{q_0 \omega}{c} \sin{(\frac{2\pi}{\lambda}(\frac{d}{2}-z)+\omega t_r)}
\end{equation}
In order to obtain (\ref{eqn:charge_dist}), integration with respect to $t_r$ needs to be performed and as a result initial condition appears. However, the effect of this initial condition on electrodynamic force fields turns out to be negligible and can be ignored. Now we can find the scalar potential due to linear (one-dimensional) charge distribution
\begin{equation}\label{eqn:scalar_potential3}
V = \frac{1}{4\pi\epsilon_0} \int_{-\frac{d}{2}}^{\frac{d}{2}} \frac{\xi(\mathbf{r}',t_r)}{\lrcurs} dz = -\frac{ q_0 \omega}{4\pi \epsilon_0 c} \int_{-\frac{d}{2}}^{\frac{d}{2}} \frac{\sin(\frac{2\pi}{\lambda}(\frac{d}{2}-z)+\omega(t-\frac{\lrcurs}{c}))}{\lrcurs} dz
\end{equation}
Integral term in (\ref{eqn:scalar_potential3}) is similar to the integral in (\ref{eqn:vector_potential}) and (\ref{eqn:sine}). Therefore, we can utilize results already derived for the vector potential. Now the scalar potential due to distributed charges along the wire becomes
\begin{equation}\label{eqn:scalar_potential4}
\begin{aligned}
& V(\mathbf{r},t) \approx -\frac{2q_0}{4\pi \epsilon_0 r}  \Big(\cos(\omega(t-\frac{r}{c})) (\frac{1}{1+\cos{\theta}} \sin(\frac{\omega d}{4c}(3+\cos{\theta})) \sin(\frac{\omega d}{4c}(1+\cos{\theta}))+\\
& \frac{1}{1-\cos{\theta}} \sin(\frac{\omega d}{4c}(1+\cos{\theta})) \sin(\frac{\omega d}{4c}(1-\cos{\theta}))) + \\
& \sin(\omega(t-\frac{r}{c})) (\frac{1}{1+\cos{\theta}} \cos(\frac{\omega d}{4c}(3+\cos{\theta})) \sin(\frac{\omega d}{4c}(1+\cos{\theta}))+\\
& \frac{1}{1-\cos{\theta}} \cos(\frac{\omega d}{4c}(1+\cos{\theta})) \sin(\frac{\omega d}{4c}(1-\cos{\theta}))) \Big)
\end{aligned}
\end{equation}
The corresponding gradient of the scalar potential is
\begin{equation}\label{eqn:time_V}
\begin{aligned}
& \nabla V (\mathbf{r},t) = \frac{2q_0 \omega}{4\pi \epsilon_0 r c} \Big(\cos(\omega(t-\frac{r}{c})) (\frac{1}{1+\cos{\theta}} \cos(\frac{\omega d}{4c}(3+\cos{\theta})) \sin(\frac{\omega d}{4c}(1+\cos{\theta}))+\\
& \frac{1}{1-\cos{\theta}} \cos(\frac{\omega d}{4c}(1+\cos{\theta})) \sin(\frac{\omega d}{4c}(1-\cos{\theta}))) - \\
& \sin(\omega(t-\frac{r}{c})) (\frac{1}{1+\cos{\theta}} \sin(\frac{\omega d}{4c}(3+\cos{\theta})) \sin(\frac{\omega d}{4c}(1+\cos{\theta}))+\\
& \frac{1}{1-\cos{\theta}} \sin(\frac{\omega d}{4c}(1+\cos{\theta})) \sin(\frac{\omega d}{4c}(1-\cos{\theta}))) \Big) \mathbf{\hat{r}}
\end{aligned}
\end{equation}

Finally we can combine (\ref{eqn:grad_v2}), (\ref{eqn:time_A}) and (\ref{eqn:time_V}) to find the electric field as $\mathbf{E}(\mathbf{r},t) = -\nabla V-\frac{\partial{\mathbf{A}}}{\partial{t}}$.
Next we need to find the magnetic field $\mathbf{B}$. However before finding the magnetic field, let us derive equation for the curl of $\mathbf{A}=A_r \mathbf{\hat{r}} + A_{\theta} \bm{\hat{\theta}}$ given in (\ref{eqn:vector_potential2})  
\begin{equation}\label{eqn:curl_A}
\nabla \times \mathbf{A} = \frac{1}{r}(\frac{\partial}{\partial{r}}(r A_{\theta}) - \frac{\partial{A_r}}{\partial{\theta}})\bm{\hat{\phi}} = (\frac{A_{\theta}}{r}+\frac{\partial{A_{\theta}}}{\partial{r}} - \frac{1}{r}\frac{\partial{A_r}}{\partial{\theta}}) \bm{\hat{\phi}};
\end{equation}
Here again, as is the case with the scalar potential, curl of the vector potential has only one component along $\bm{\hat{\phi}}$, because the vector potential does not depend on $\phi$ or $\bm{\hat{\phi}}$. If we use $\mathbf{\hat{z}} = \cos{\theta}\mathbf{\hat{r}}-\sin{\theta}\bm{\hat{\theta}}$ in (\ref{eqn:vector_potential2}), then substitute it to (\ref{eqn:curl_A}) and apply the assumptions (\ref{eqn:as1}) and (\ref{eqn:as3}), we get after somewhat lengthy derivation the magnetic field as $\mathbf{B} = \nabla \times \mathbf{A}$.
At last we found electric (\ref{eqn:electric_r_theta}) and magnetic (\ref{eqn:magnetic_phi}) fields radiated by the electric dipole under the assumptions (\ref{eqn:as1})-(\ref{eqn:as3}).

\section{Results}

In effect, the electric field of the dipole can be written as $\mathbf{E} = E_r \mathbf{\hat{r}} + E_{\theta} \bm{\hat{\theta}}$ and it consists of radial $E_r$ and transverse $E_{\theta}$ components. If simple trigonometric identities are used, then we can write the radial and transverse components of the electric field in a form more suitable for later analysis
\begin{subequations}\label{eqn:electric_r_theta}
	\begin{equation}\label{eqn:electric_r}
	\begin{aligned}
	& E_r = -\frac{q_0}{4\pi \epsilon_0} \frac{\omega}{r c} \Bigg( \cos(\omega(t-\frac{r}{c})) \\ 
	& \Big( \sin(\frac{\omega d}{2c}\cos{\theta}) + \sin(\frac{\omega d}{2c}) + \frac{1-\cos{\theta}}{1+\cos{\theta}}(\sin(\frac{\omega d}{2c}(2+\cos{\theta}))-\sin(\frac{\omega d}{2c}))\Big) + \\
	& \sin(\omega(t-\frac{r}{c})) \Big( \cos(\frac{\omega d}{2c})-\cos(\frac{\omega d}{2c}\cos{\theta}) + \frac{1-\cos{\theta}}{1+\cos{\theta}}(\cos(\frac{\omega d}{2c}(2+\cos{\theta}))-\cos(\frac{\omega d}{2c})) \Big) \Bigg)
	\end{aligned}	
	\end{equation}
	\begin{equation}\label{eqn:electric_theta}
	\begin{aligned}
	& E_{\theta} = -\frac{q_0}{4\pi \epsilon_0} \frac{\omega}{r c} \Bigg( \cos(\omega(t-\frac{r}{c})) \\ 
	& \Big( \frac{\sin{\theta}}{1+\cos{\theta}} (\sin(\frac{\omega d}{2c}(2+\cos{\theta}))-\sin(\frac{\omega d}{2c})) +  \frac{\sin{\theta}}{1-\cos{\theta}}(\sin(\frac{\omega d}{2c})-\sin(\frac{\omega d}{2c}\cos{\theta}))\Big) + \\
	& \sin(\omega(t-\frac{r}{c})) \Big( \frac{\sin{\theta}}{1+\cos{\theta}} (\cos(\frac{\omega d}{2c}(2+\cos{\theta}))-\cos(\frac{\omega d}{2c})) +  \frac{\sin{\theta}}{1-\cos{\theta}}(\cos(\frac{\omega d}{2c})-\cos(\frac{\omega d}{2c}\cos{\theta}))\Big) \Bigg)
	\end{aligned}	
	\end{equation}
\end{subequations}
The magnetic field can be written as $\mathbf{B} = B_{\phi} \bm{\hat{\phi}}$, where again after applying some trigonometric identities
\begin{equation}\label{eqn:magnetic_phi}
\begin{aligned}
& B_{\phi} = -\frac{q_0}{4\pi \epsilon_0} \frac{\omega}{r c^2} \Bigg( \cos(\omega(t-\frac{r}{c})) \\ 
& \Big( \frac{\sin{\theta}}{1+\cos{\theta}} (\sin(\frac{\omega d}{2c}(2+\cos{\theta}))-\sin(\frac{\omega d}{2c})) +  \frac{\sin{\theta}}{1-\cos{\theta}}(\sin(\frac{\omega d}{2c})-\sin(\frac{\omega d}{2c}\cos{\theta}))\Big) + \\
& \sin(\omega(t-\frac{r}{c})) \Big( \frac{\sin{\theta}}{1+\cos{\theta}} (\cos(\frac{\omega d}{2c}(2+\cos{\theta}))-\cos(\frac{\omega d}{2c})) +  \frac{\sin{\theta}}{1-\cos{\theta}}(\cos(\frac{\omega d}{2c})-\cos(\frac{\omega d}{2c}\cos{\theta}))\Big) \Bigg)
\end{aligned}	
\end{equation}                                                     

Let us try to qualitatively understand the obtained results. The first observation is that electric (\ref{eqn:electric_r_theta}) and magnetic (\ref{eqn:magnetic_phi}) fields are inversely proportional to the distance $r$ in the first power. Therefore we conclude that these are indeed far fields (long-range) radiated by the electric dipole. The second observation is that if we compare (\ref{eqn:electric_theta}) and (\ref{eqn:magnetic_phi}), then $B_{\phi} = \frac{{E}_{\theta}}{c}$, $B_{\phi}$ and $E_{\theta}$ are in phase, they are mutually perpendicular and they are perpendicular to the travel direction $\mathbf{\hat{r}}$. Thus we conclude that $B_{\phi}$ and $E_{\theta}$ together comprise \emph{transverse electromagnetic} component of the radiation. The reason behind our logic is that longitudinal electric field component does not contribute anything to the Poynting vector and so can be disregarded while considering transverse electromagnetic radiation. On the other hand, $E_r$ is directed radially outward along the travel direction $\mathbf{\hat{r}}$ and there is no magnetic field which can be associated (linked) with this electric field.  Therefore we conclude that $E_r$ is pure \emph{radial} or \emph{longitudinal electric} component of the radiation. Also we notice that total electric field lies in $r-\theta$ plane, while total magnetic field is perpendicular to that plane (along $\bm{\hat{\phi}}$). Thus total electric field is still orthogonal to the total magnetic field. The third observation is that, apart from terms contained within large brackets, $E_r$ and $E_{\theta}$ are equal in magnitude. Thus it suggests that longitudinal electric field is on par in terms of intensity with its traditional transverse counterpart, i.e. it is not some second order ``faint'' effect. In other words, longitudinal electric field is as real as the transverse electric field. We can call $E_r$ pure ``electric wave'', in contrast to electromagnetic wave, because it is electric field radiated and traveling outward by the dipole (without accompanying magnetic component). 

Another important item that we should point out is that during the derivation of these results up to this point we actually never used explicitly the assumption (\ref{eqn:as2}). We used it implicitly by carrying over (not reducing) all terms which contain $\frac{c}{\omega d}$, for example, you see terms with it in (\ref{eqn:electric_r_theta}) and (\ref{eqn:magnetic_phi}). Therefore we can say that the above derivation was done using only two assumptions (\ref{eqn:as1}) and (\ref{eqn:as3}). However we know that during the derivation of radiation in the standard classical case, the inverse assumption (\ref{eqn:as_old}) (or $\frac{\omega d} {c}\ll 1$) is used in addition to (\ref{eqn:as1}) and (\ref{eqn:as3}). Thus, judging only by the number of assumptions used during the derivation, we can conclude that the derivation presented here is more general than the traditional one. However that also gives us opportunity to check our solution, because with the assumption (\ref{eqn:as_old}) our solution should be reduced to the traditional classical electric dipole radiation results. Let us verify that, using the following approximations obtained by employing (\ref{eqn:as_old}) 
\begin{subequations}\label{eqn:old_clas_as}
	\begin{equation}
	\sin(\frac{\omega d} {c}) \approx \frac{\omega d} {c}
	\end{equation}
	\begin{equation}
	\cos(\frac{\omega d} {c}) \approx 1
	\end{equation}
\end{subequations}
By applying (\ref{eqn:old_clas_as}) to (\ref{eqn:electric_r_theta}) and (\ref{eqn:magnetic_phi}), the following is obtained
\begin{subequations}
	\begin{equation}
	\begin{aligned}
	& E_r \approx -\frac{q_0}{4\pi \epsilon_0} \frac{\omega}{r c} \Bigg( \cos(\omega(t-\frac{r}{c})) \Big( \frac{\omega d}{2c}\cos{\theta} + \frac{\omega d}{2c} + \frac{1-\cos{\theta}}{1+\cos{\theta}}(\frac{\omega d}{2c}(2+\cos{\theta})-\frac{\omega d}{2c})\Big) + \\
	& \sin(\omega(t-\frac{r}{c})) \Big( 1-1 + \frac{1-\cos{\theta}}{1+\cos{\theta}}(1-1) \Big) \Bigg) = -\frac{q_0}{4\pi \epsilon_0} \frac{\omega^2 d}{r c^2} \cos(\omega(t-\frac{r}{c}))
	\end{aligned}
	\end{equation}
	\begin{equation}
	\begin{aligned}
	& E_{\theta} \approx -\frac{q_0}{4\pi \epsilon_0} \frac{\omega}{r c} \Bigg( \cos(\omega(t-\frac{r}{c})) \Big( \frac{\sin{\theta}}{1+\cos{\theta}} (\frac{\omega d}{2c}(2+\cos{\theta})-\frac{\omega d}{2c}) +  \frac{\sin{\theta}}{1-\cos{\theta}}(\frac{\omega d}{2c}-\frac{\omega d}{2c}\cos{\theta})\Big) + \\
	& \sin(\omega(t-\frac{r}{c})) \Big( \frac{\sin{\theta}}{1+\cos{\theta}} (1-1) + \frac{\sin{\theta}}{1-\cos{\theta}}(1-1)\Big) \Bigg) = -\frac{q_0}{4\pi \epsilon_0} \frac{\omega^2 d}{r c^2} \cos(\omega(t-\frac{r}{c})) \sin{\theta}
	\end{aligned}
	\end{equation}
	\begin{equation}
	\begin{aligned}
	& B_{\phi} \approx -\frac{q_0}{4\pi \epsilon_0} \frac{\omega}{r c^2} \Bigg( \cos(\omega(t-\frac{r}{c})) \Big( \frac{\sin{\theta}}{1+\cos{\theta}} (\frac{\omega d}{2c}(2+\cos{\theta})-\frac{\omega d}{2c}) +  \frac{\sin{\theta}}{1-\cos{\theta}}(\frac{\omega d}{2c}-\frac{\omega d}{2c}\cos{\theta})\Big) + \\
	& \sin(\omega(t-\frac{r}{c})) \Big( \frac{\sin{\theta}}{1+\cos{\theta}} (1-1) + \frac{\sin{\theta}}{1-\cos{\theta}}(1-1)\Big) \Bigg) = -\frac{q_0}{4\pi \epsilon_0} \frac{\omega^2 d}{r c^3} \cos(\omega(t-\frac{r}{c})) \sin{\theta}
	\end{aligned}
	\end{equation}
\end{subequations}
Thus we can clearly see that transverse component of our solution under the assumption (\ref{eqn:old_clas_as}) reduces to the standard classical result. However, contrary to the classical result, the longitudinal component is nonzero. This is due to the linearly distributed charges $\xi$ along the length of the wire, which are not taken into account during the derivation of the classical dipole radiation. 

\begin{figure}
	\centering
	\includegraphics[width=0.6\columnwidth]{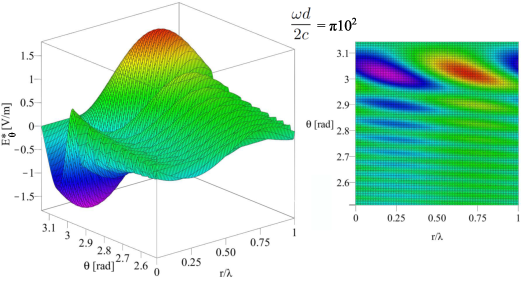}
	\caption{Transverse electric field as a function of $r$ and $\theta$. 
		For clarity the scaled electric field $E_{\theta}^* = E_{\theta}/10$ (reduced by ten) is plotted.}
	\label{fig:electrictran}
\end{figure}

Let us try to visualize the radiation pattern for the new case. We assume the following parameters: $q_0 = 10^{-9}$ Q, $d = 1$ m, $\lambda = 10^{-2}$ m, $r=r_0=10^4$ m, $t=1$ s, $\omega = \frac{2\pi c}{\lambda} = 6\cdot 10^{10} \pi$ rad/s (microwave frequency). Note that assumed values for $r$, $d$ and $\lambda$ satisfy our assumptions (\ref{eqn:as1})-(\ref{eqn:as3}). With the given values we want to plot longitudinal (\ref{eqn:electric_r}) and transverse (\ref{eqn:electric_theta}) electric field values as a function of $\theta$ and $r$ in the following range $0 \leq \theta \leq \pi$ and $r_0 \leq r \leq r_0+\lambda$. Interestingly, when these functions were plotted we found that longitudinal and transverse electric fields are zero for all values of $\theta$ except at the angles close to $\theta=\pi$ (and $\theta=0$ for transverse field). 
Therefore, in Fig. \ref{fig:electriclong} and in \ref{fig:electrictran}, for clarity, longitudinal and transverse electric fields are shown for a small region  $\frac{4\pi}{5} \leq \theta \leq \pi$. We can also say that these figures depict ``time shots'' of electric fields as they are traveling in space at $t=1$ s. A range of the dimensionless parameter $0 \leq \frac{r}{\lambda} \leq 1$ corresponds to $r_0 \leq r \leq r_0+\lambda$. In order to save space with numbering of the $z$ axis in the plots, electric fields had to be scaled as $E_r^* = E_r/200$ for longitudinal electric field and as $E_{\theta}^* = E_{\theta}/10$ for transverse electric field. 
For $r<r_0$ and $r>r_0$ electric fields are just cyclic repetitions of these plots (this is valid only if variation of $r$ is small enough to neglect amplitude variation of the electric fields due to the inverse $r$ term). By comparing Fig. \ref{fig:electriclong} and \ref{fig:electrictran} we notice the following (let us remind that this is valid for the chosen values of the parameters):

\begin{figure}
	\centering
	\includegraphics[width=0.6\columnwidth]{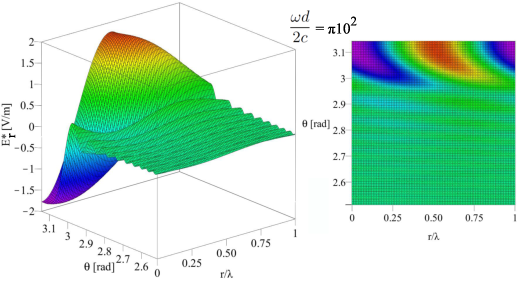}
	\caption{Longitudinal electric field as a function of $r$ and $\theta$. 
		For clarity the scaled electric field $E_r^* = E_r/200$ (reduced by two hundred) is plotted.}
	\label{fig:electriclong}
\end{figure}

\begin{itemize}
	\item The magnitude of the longitudinal electric far field is larger by an order than the magnitude of the transverse electric far field.
	\item Maximum and minimum values of the longitudinal electric wave lie exactly on $\theta=\pi$, while max and min values of the transverse electric field lie not exactly but slightly off the angles $\theta=0$ and $\theta=\pi$. Transverse electric field is actually close to zero at $\theta=0$ and $\theta=\pi$ values.
	\item Radiation of the longitudinal and transverse waves happens in very narrow cone beam with planar angle around $\triangle \theta \approx 0.2$ rad.
	\item Shape of the longitudinal electric field wavelet is different than the shape of the transverse field wavelet.
	\item Spatially, min values of the longitudinal field roughly correspond to min values of the transverse field, and vice-versa.

\end{itemize}

The next step for us is to find the total radiated power. Longitudinal and transverse electric waves carry energy away from the dipole. Let us first evaluate the average net power radiated by the longitudinal electric field. Due to the fact that Poynting vector accounts only for energy transfer by the transverse electric and magnetic fields, we have to use longitudinal electric field energy density in order to calculate its energy transfer. Energy density $\eta_r$ of the longitudinal electric field is given as 
\begin{equation}\label{eqn:energydensity_long}
\eta_r = \frac{1}{2}\epsilon_0 E_r^2
\end{equation}
If we assume that longitudinal electric field also travels outward with the speed $c$, then magnitude $W_r$ of energy transferred away by the longitudinal waves per unit time $t$ per unit area $S$ is given as
\begin{equation}\label{eqn:powerperarea_long}
\frac{d\mathbf{W}_r}{dt dS} = \eta_r c \mathbf{\hat{r}} = \frac{1}{2}\epsilon_0 E_r^2 c \mathbf{\hat{r}}
\end{equation}
here $\mathbf{\hat{r}}$ just shows direction of energy flow. Average energy per period passing over unit area by the longitudinal waves, which is equal to average power per area, can be found as:
\begin{equation}\label{eqn:averagepower_long}
\frac{d\bar{\mathbf{W}}_r}{dS} = \frac{1}{2}\epsilon_0 c \mathbf{\hat{r}} \frac{1}{T} \int_{0}^{T} E^2_r(t) dt
\end{equation}
In order to evaluate the above integral we can rewrite (\ref{eqn:electric_r})
\begin{equation}\label{eqn:electric_r2}
\begin{aligned}
& E_r(t) = -\frac{q_0}{4\pi \epsilon_0} \frac{\omega}{r c} \Bigg( \cos(\omega(t-\frac{r}{c})) \mathcal{A} + \sin(\omega(t-\frac{r}{c})) \mathcal{B} \Bigg)
\end{aligned}	
\end{equation}
where $\mathcal{A} = \sin(\frac{\omega d}{2c}\cos{\theta}) + \sin(\frac{\omega d}{2c}) + \frac{1-\cos{\theta}}{1+\cos{\theta}}(\sin(\frac{\omega d}{2c}(2+\cos{\theta}))-\sin(\frac{\omega d}{2c}))$
and $\mathcal{B} = \cos(\frac{\omega d}{2c})-\cos(\frac{\omega d}{2c}\cos{\theta}) + \frac{1-\cos{\theta}}{1+\cos{\theta}}(\cos(\frac{\omega d}{2c}(2+\cos{\theta}))-\cos(\frac{\omega d}{2c}))$ do not depend on time. If (\ref{eqn:electric_r2}) into (\ref{eqn:averagepower_long}) are substituted, and the following identities
\begin{subequations}
	\begin{equation}
	\frac{1}{T} \int_{0}^{T} \cos^2(\omega(t-\frac{r}{c})) dt = \frac{1}{2}
	\end{equation}
	\begin{equation}
	\frac{1}{T} \int_{0}^{T} \sin^2(\omega(t-\frac{r}{c})) dt = \frac{1}{2}
	\end{equation}
	\begin{equation}
	\frac{1}{T} \int_{0}^{T} \cos(\omega(t-\frac{r}{c})) \sin(\omega(t-\frac{r}{c})) dt = 0
	\end{equation}
\end{subequations}
are used, then we obtain
\begin{equation}\label{eqn:pattern_long}
\frac{d\bar{\mathbf{W}}_r}{dS} = \frac{1}{4} \epsilon_0 c (\frac{q_0}{4\pi \epsilon_0} \frac{\omega}{r c})^2 (\mathcal{A}^2+\mathcal{B}^2) \mathbf{\hat{r}}
\end{equation}
Infinitesimal surface area in spherical coordinate system at constant radius $r$ is given as $d\mathbf{S} = r^2 \sin{\theta} d\theta d\phi \mathbf{\hat{r}}$. The net average power radiated by the longitudinal electric waves of the dipole is 
\begin{equation}\label{eqn:totalpower_long}
\begin{aligned}
& \bar{W}_r = \int \frac{d\bar{\mathbf{W}}_r}{dS} \cdot d\mathbf{S} = \frac{1}{4} \epsilon_0 c (\frac{q_0}{4\pi \epsilon_0} \frac{\omega}{r c})^2 \int\limits_{0}^{2\pi} \int_{0}^{\pi} (\mathcal{A}^2+\mathcal{B}^2) r^2 \sin{\theta} d\theta d\phi = \\
& \frac{q_0^2 \omega^2}{32 \pi \epsilon_0 c} \int_{0}^{\pi} (\mathcal{A}^2+\mathcal{B}^2) \sin{\theta} d\theta
\end{aligned}
\end{equation}
As we can see the average radiated longitudinal power of the dipole does not depend on $r$, which is expected. 

With the same analysis we can obtain the expression for average net power, radiated by the transverse electric and magnetic waves of the dipole. For that the Poynting vector is given as
\begin{equation}\label{eqn:powerperarea_tran}
\mathbf{P} = \frac{1}{\mu_0} E_{\theta} \bm{\hat{\theta}} \times B_{\phi} \bm{\hat{\phi}} = \epsilon_0 E^2_{\theta} c \mathbf{\hat{r}}
\end{equation}
The average energy per period per area radiated away can be found
\begin{equation}\label{eqn:averagepower_tran}
\bar{\mathbf{P}} = \epsilon_0 c \mathbf{\hat{r}} \frac{1}{T}  \int_{0}^{T} E^2_{\theta}(t) dt
\end{equation}
As before, in order to evaluate (\ref{eqn:averagepower_tran}), (\ref{eqn:electric_theta}) can be rewritten as
\begin{equation}\label{eqn:electric_theta2}
\begin{aligned}
& E_{\theta} = -\frac{q_0}{4\pi \epsilon_0} \frac{\omega}{r c} \Bigg( \cos(\omega(t-\frac{r}{c})) \mathcal{C} + \sin(\omega(t-\frac{r}{c})) \mathcal{D} \Bigg)
\end{aligned}	
\end{equation}
where $\mathcal{C} = \frac{\sin{\theta}}{1+\cos{\theta}} (\sin(\frac{\omega d}{2c}(2+\cos{\theta}))-\sin(\frac{\omega d}{2c})) +  \frac{\sin{\theta}}{1-\cos{\theta}}(\sin(\frac{\omega d}{2c})-\sin(\frac{\omega d}{2c}\cos{\theta})) $ and $\mathcal{D} = \frac{\sin{\theta}}{1+\cos{\theta}} (\cos(\frac{\omega d}{2c}(2+\cos{\theta}))-\cos(\frac{\omega d}{2c})) +  \frac{\sin{\theta}}{1-\cos{\theta}}(\cos(\frac{\omega d}{2c})-\cos(\frac{\omega d}{2c}\cos{\theta}))$. Again $\mathcal{C}$ and $\mathcal{D}$ do not depend on time. Therefore for (\ref{eqn:averagepower_tran}) we obtain
\begin{equation}\label{eqn:pattern_tran}
\bar{\mathbf{P}}  = \frac{1}{2} \epsilon_0 c (\frac{q_0}{4\pi \epsilon_0} \frac{\omega}{r c})^2 (\mathcal{C}^2+\mathcal{D}^2) \mathbf{\hat{r}}
\end{equation}
In order to find the average net power radiated by the transverse electric and magnetic waves, the above expression needs to be integrated with respect to area
\begin{equation}\label{eqn:totalpower_tran}
\begin{aligned}
& \bar{W}_t = \int \bar{\mathbf{P}} \cdot d\mathbf{S} =  \frac{1}{2} \epsilon_0 c (\frac{q_0}{4\pi \epsilon_0} \frac{\omega}{r c})^2 \int\limits_{0}^{2\pi} \int_{0}^{\pi} (\mathcal{C}^2+\mathcal{D}^2) r^2 \sin{\theta} d\theta d\phi = \\
& \frac{q_0^2 \omega^2}{16 \pi \epsilon_0 c} \int_{0}^{\pi} (\mathcal{C}^2+\mathcal{D}^2) \sin{\theta} d\theta
\end{aligned}
\end{equation}
The average total power (energy per period) radiated by the electric dipole is equal to the sum of net longitudinal (\ref{eqn:totalpower_long}) and transverse (\ref{eqn:totalpower_tran}) powers
\begin{equation}\label{eqn:total_power}
\bar{W}_{total} = \bar{W}_r + \bar{W}_t
\end{equation}
We could not find the analytical solution for the integrals in (\ref{eqn:totalpower_long}) and (\ref{eqn:totalpower_tran}). Therefore we are providing numerical results. Just for the sake of comparison, we present the classical result for total power (transverse only) radiated by the dipole with the assumption (\ref{eqn:as_old})
\begin{equation}\label{eqn:totalpower_old}
\bar{W}_{old} = \frac{q_0^2 d^2 \omega^4}{12 \pi \epsilon_0 c^3} 
\end{equation}

\begin{figure}[t]
	\centering
	\includegraphics[width=1.0\columnwidth]{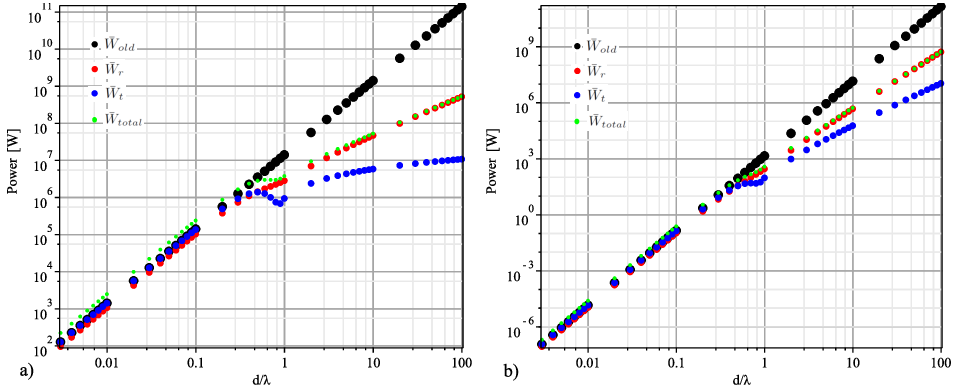}
	\caption{Average longitudinal power $\bar{W}_r$ (red dot), average transverse power $\bar{W}_t$ (blue dot), average total power $\bar{W}_{total}$ (green dot) and total power according to classical solution $\bar{W}_{old}$ (black dot) radiated by the dipole as a function of $\frac{d}{\lambda}$. a) for variation of $d$ ($\lambda = 10^{-2}$ m), b) for variation of $\lambda$ ($d=1$ m).}
	\label{fig:powers}
\end{figure}

In order to see the effect of the ratio $\frac{d}{\lambda}$ on power, we fixed $\lambda = 10^{-2}$ m and varied $d$. For the dipole with charge separation $d$, we can define ``\emph{fundamental wavelength}'' and ``\emph{fundamental frequency}'' as $\lambda_0=d$ and $\omega_0 = \frac{2\pi c}{\lambda_0}$. The longitudinal power (\ref{eqn:totalpower_long}), transverse power (\ref{eqn:totalpower_tran}), total power (\ref{eqn:total_power}) and total power according to classical solution (\ref{eqn:totalpower_old}) radiated by the dipole as a function of $\frac{d}{\lambda}$ are shown in Fig. \ref{fig:powers}a. As we can see, total power according to the classical solution, which is the transverse power, has constant slope in the log scale plot. For $\frac{d}{\lambda} \ll 1$, which in another words means for $\omega \ll \omega_0$,  total power (\ref{eqn:total_power}) consists of both longitudinal (\ref{eqn:totalpower_long}) and transverse (\ref{eqn:totalpower_tran}) powers. Longitudinal power is roughly equal to three quarters of the transverse power. This is in contrast to the classical solution, where total power is dominated by the transverse power. 
In this region due to assumption (\ref{eqn:as_old}) classical solution is valid, and so $\bar{W}_{total} \sim d^2$. For $\frac{d}{\lambda} \gg 1$, or in other words for $\omega \gg \omega_0$, total power consists of mostly longitudinal power, which is several orders of magnitude larger than transverse power. This means that in this region transverse power is negligible. Total power from classical solution is several orders of magnitude larger than total power (\ref{eqn:total_power}). In this region due to assumption (\ref{eqn:as_old}) classical solution is not valid, and from the figure it can be found that $\bar{W}_{total} \sim d$. 
It is interesting to note that both transverse and longitudinal powers have some dip around $\omega \sim \omega_0$, which we speculate would be interesting to observe experimentally. 

\begin{figure}[t]
	\centering
	\includegraphics[width=0.6\columnwidth,angle=-90]{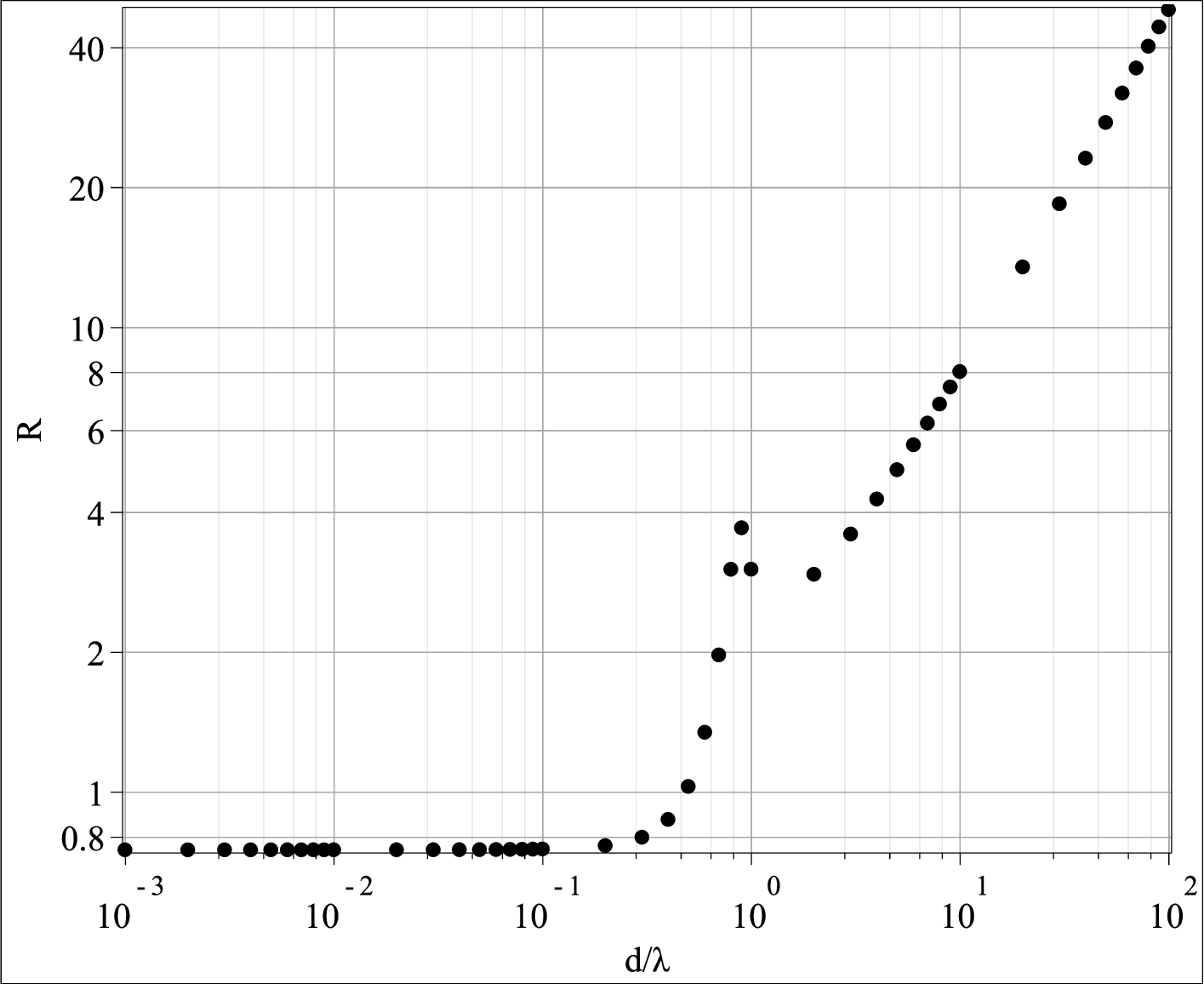}
	\caption{Ratio of longitudinal power and transverse power radiated by the dipole as a function of $\frac{d}{\lambda}$.}
	\label{fig:ratio_power}
\end{figure}

In order to see the effect of $\lambda$ (or of $\omega$) on the radiated power, we decided to plot the power again as a function of $\frac{d}{\lambda}$, but in this case by fixing the $d=1$ m and by varying instead the $\lambda$, Fig. \ref{fig:powers}b. Generally it conveys the same information as in Fig. \ref{fig:powers}a. However it can be noticed that for $\frac{d}{\lambda} \ll 1$ $\bar{W}_{total} \sim \omega^4$, which is the same as classical result $\bar{W}_{old}$. While for $\frac{d}{\lambda} \gg 1$ $\bar{W}_{total} \sim \omega^3$. Thus, it can be concluded that for $\frac{d}{\lambda} \ll 1$ $\bar{W}_{total} \sim d^2 \omega^4$, while for $\frac{d}{\lambda} \gg 1$ $\bar{W}_{total} \sim d\omega^3$. Ratio of longitudinal to transverse power
\begin{equation}
R = \frac{\bar{W}_r}{\bar{W}_t}
\end{equation}
is shown in Fig. \ref{fig:ratio_power}. This figure confirms our previous observations, for $\frac{d}{\lambda} \ll 1$ longitudinal power is on par with the transverse (longitudinal power is equal to three quarters of the transverse power), while for $\frac{d}{\lambda} \gg 1$ transverse power is negligible.

In order to see the radiation pattern plot of the longitudinal electric waves of the dipole system, the expression (\ref{eqn:pattern_long}) should be drawn as a function of $\theta$. While plotting the expression (\ref{eqn:pattern_tran}) would provide radiation pattern for transverse electric or magnetic fields of the dipole. 

\section{Discussion}\label{sec:discussion}

We can summarize the results as the following. For $0 \leq \omega \ll \omega_0$, where $\omega_0$ is the fundamental frequency of an electric dipole as defined above, the total radiated power coincides with the classical solution, while transverse power is on par with longitudinal power. Classical solution is valid in this region and total radiated power is proportional to the fourth degree of the frequency and to the second degree of the charge separation distance. For $\omega \gg \omega_0$ the longitudinal power is dominant and the classical solution greatly overestimates the total radiated power. In this region classical solution is invalid and total radiated power is proportional to the third degree of the frequency and to the charge separation distance. Thus classical solution predicts correctly only the transverse radiated power and only in the $0 \leq \omega \ll \omega_0$ region. In the region $\omega \sim \omega_0$ both longitudinal and transverse powers are equal. It was shown that under the assumption $d \ll \lambda$: a) classical solution correctly describes the transverse electromagnetic waves only; b) longitudinal electric waves are non-negligible; c) total radiated power is proportional to the fourth degree of the frequency and to the second degree of the charge separation distance; d) our transverse solution reduces to the classical solution. In case $\lambda \ll d$: a) the classical solution is not valid and it overestimates the total radiated power; b) longitudinal electric waves are dominant and transverse electromagnetic waves are negligible; c) total radiated power is proportional to the third degree of the frequency and to the charge separation distance; d) most of the power is emitted in a narrow beam along the dipole axis, thus emission of waves is focused as with lasers. Longitudinal electric fields are commonly derived for small electric dipoles~\cite{heald1995classical,balanis2005antenna}. However, due to $1/r^2$ dependence, their magnitude drops off quickly (faster than its transverse component) as distance is increased. Therefore, it is commonly accepted that longitudinal electric fields have significance only in near field around the dipole. Our results challenge this point of view.

From the fundamental principles, in order to analyze any given electrodynamic system, firstly, the electrodynamic potentials ($V$ and $\mathbf{A}$) of the system need to be calculated. Then the electrodynamic forces ($\mathbf{E}$ and $\mathbf{B}$) of the system can be found from electrodynamic potentials. For discrete systems the electrodynamic scalar potential is evaluated by summation, while for continuous systems it is evaluated by integration over source coordinates of potentials due to distributed charges. The vector potential due to current sources is evaluated in the same way. Then in order to find electrodynamic forces, we need to differentiate electrodynamic potentials with respect to observation point coordinates. For static systems, the resultant potential and force fields are functions of space coordinates only. While for dynamic systems these fields are functions of space and time coordinates. From mathematical point of view, evaluation of the potentials and forces can be done in two ways: I) by direct summation and integration, or II) by power series expansion around arbitrary point in space (for static systems) and in time (for dynamic systems). In the second option, Taylor series expansion is applied for numerator (source terms) and for denominator (distance) terms of the potentials \cite{zangwill2013modern,heald1995classical}. Of course, in this case, the potentials can be evaluated only if the series converges. In order for the power series to converge, the assumption that $d\ll r$ is utilized for the static system and the assumptions $d \ll \lambda \ll r$ are utilized for the dynamic dipole systems. As a result of these assumptions, potentials can be evaluated through power series expansion, and each term of the resultant expansion can be characterized as electric dipole moments of diverse order (dipole, quadrupole, etc.). However it does not mean that these assumptions are necessary for evaluation of potentials through direct method or through other Taylor series expansions. In this paper, the Taylor series expansion was utilized only for the $r$ term in the potentials, while $\lambda$ and $d$ terms were integrated fully in order to find the potentials and forces under the assumption $\lambda \ll d \ll r$. 

In some books derivation of radiation of the classical electric dipole is done for the case when point charges have constant charge and are oscillating \cite{WalterGreiner}. While in other derivations point charges are taken as fixed in space, but their charge magnitudes are oscillating \cite{DavidGriffiths}. For the classical dipole case ($d \ll \lambda$) these two approaches lead to the same results. In this work we extended the latter case, while similar extension of the former case is impossible. This comes from the fact that point charges, while oscillating with period $T$, can not move faster than $c$
\begin{equation}
\frac{v}{c} \approx \frac{d}{Tc} =\frac{d}{\lambda}
\end{equation}
Therefore extension of the case, where point charges of the dipole have constant charge but are moving (oscillating), is impossible for $\frac{d}{\lambda} \gg 1$. Even though in the region $\frac{d}{\lambda} \ll 1$ the above two approaches are equivalent, we speculate that in the region $\frac{d}{\lambda} \sim 1$ these two approaches lead to different outcomes. 

We note that even though the results for electric and magnetic fields given in (\ref{eqn:electric_r_theta}) and (\ref{eqn:magnetic_phi}) are generalized cases of the classical solution, the results for longitudinal electric field are strictly valid only for the case $\frac{d}{\lambda} = N \in \mathbb{N}$, where $N$ is non-negative integer ($N=0,1,2,...,\infty$). This is due to two factors: the first is the specific form of scalar potential taken in (\ref{eqn:potential}), while the second is due to distributed charges (\ref{eqn:charge_dist}). The net integral of distributed charges along the antenna length is zero for $\frac{d}{\lambda} = N$, thus only for this case the total charge is conserved. As a result, our results for longitudinal electric field component in the region $\frac{d}{\lambda} \ll 1$ might be incorrect. However, results for longitudinal electric field in the region $\frac{d}{\lambda} \gg 1$ are correct. The results for transverse electric and magnetic fields are valid for general case $\frac{d}{\lambda} = R \in \mathbb{R}$, where $R$ is positive real number. 

After obtaining these results, let us now be bold by turning on our imagination and let us discuss some issues or questions which the new findings open up. 

We know that transverse electromagnetic waves are transmitted by photons. Are there longitudinal photons, which transmit the longitudinal pure electric waves? We speculate that if these longitudinal photons exist, then they might have properties completely different than transverse photons.

For the parameters that were used to obtain the results, 
the radiation pattern is very narrowly focused beam, like a laser beam. We speculate that the larger is the ratio $\frac{d}{\lambda}$, the more focused the beam is. Focusing of the beam is achieved just by increasing the separation distance $d$, without using any reflecting mirrors. The larger is the ratio $\frac{d}{\lambda}$, the more power is radiated by longitudinal waves. Therefore it will be interesting to investigate the possible usage of the longitudinal electric waves for communication and possibly for power transmission. 

We speculate that longitudinal electric waves can shed a light and explain the so called EM drive phenomena \cite{white2017measurement}. Also phenomena of displacement current, which is used to describe the magnetic field arising around capacitance when alternating current is passing through it, might be explained by this longitudinal electric waves. In fact, we note that classical solution suggests that point electric charges, harmonically oscillating along the straight line, emit mostly transverse waves perpendicular to the line. While the new results indicate that when $\frac{d}{\lambda} \gg 1$, point electric charges, with harmonically oscillating charge magnitude, emit longitudinal and transverse waves mostly along the line joining the two charges. Thus if we think of two capacitance plates as consisting of many infinitesimal point charges, then we believe that capacitance is the best emitter of longitudinal waves in the direction perpendicular to the plates. Vector potential waves, which we believe are longitudinal electric waves, were experimentally observed in \cite{zimmerman2013reception}. Also scalar potential waves, which again were probably longitudinal electric waves, were observed experimentally in \cite{monstein2002} and in~\cite{zaimidoroga2016electroscalar}. In the former paper the scalar potential waves were emitted by the spherical antenna. Longitudinal electric waves were observed in air \cite{glakos1991anomalous}. Longitudinal electrodynamic forces, observed in \cite{johansson1996longitudinal}, and longitudinal Ampere forces in \cite{graneau1984longitudinal} could be due to the proposed longitudinal electric waves.


Calculation of energy transfer by longitudinal electric waves in (\ref{eqn:energydensity_long})-(\ref{eqn:totalpower_long}) contradicts accepted form of the electrodynamic conservation of energy principle. Therefore, we would like to make the following two comments in defense of our approach. The first is that electrodynamic force fields ($\mathbf{B}$ and $\mathbf{E}$) do not specify the electromagnetic system uniquely~\cite{reiss2012modified,reiss2017physical}. On the contrary, electrodynamic potential fields ($\mathbf{A}$ and $V$) fully define and specify the system. As a result, reformulation of energy conservation principle in terms of electrodynamic potentials might be necessary. In other words, accepted form of the energy conservation principle in terms of electric and magnetic energy densities and the Poynting vector might not account for all of energy terms. Electromagnetic energy and power flux were reformulated using potentials and their unmixed derivatives in~\cite{puthoff2016electromagnetic}, where it was shown that various alternative expressions give the same net energy and power flow, but different energy and power flow distributions. The second comment is the existence of papers, where authors demonstrate inconsistencies in the standard electrodynamic conservation of energy principle~\cite{jeffries1992conservation,gough1982poynting,puthoff2016electromagnetic}. For example, Jeffries shows that utilization of the Poynting vector leads to paradoxes, and he states that the requirement that Poynting formula should be used as a conservation law only when integrated over the closed surface was devised in order to artificially avoid these paradoxes. Similar to Puthoff's paper, Jeffries also uses potentials and their derivatives in order to devise new energy density and power flux terms. Interestingly, analogous to our paper, Jeffries also derived that the total power radiated by an antenna using his formulation of energy conservation law leads to twice the value of classical power. There were numerous other attempts to come up with alternative energy conservation principles. Formulation of the energy conservation, which consistently reflects the notion of energy, was developed in~\cite{jeffries1992conservation}. Shortcomings of the Poynting measures were described in~\cite{slepian1942energy}. The author then derived multiple alternative energy conservation principles. Also in~\cite{hines1952electromagnetic} the author argues that alternative formulations of energy conservation principle are equally valid, as long as $\mathbf{E} \cdot \mathbf{J}$ is the only observationally relevant term. Alternative formulations of energy conservation principle for longitudinal electric waves, where no Poynting vector is present, were formulated in~\cite{zaimidoroga2016electroscalar}. Shortcomings of the Poynting vector and other open questions in electrodynamics are discussed in~\cite{ribaric1990conservation}. Therefore, it can be concluded that commonly accepted conservation of energy principle might be incomplete. We view our work as another small contribution to the discussion of unresolved questions in classical electrodynamics, and to the search of the truth.

Final note is that existence of longitudinal electric waves contradicts one of the Maxwell equations (Gauss's law) $\nabla \cdot \mathbf{E} = 0$, which in the given form is thought to be valid in vacuum for volumes of space outside of electric charges. However in our next work, which is almost complete, we will prove that Gauss law is only valid for a point charge with constant magnitude $q(t) = const$, no matter how it moves. Results show that for a system in which total charge is not a constant, but instead is a function of time $q(t)$, the Gauss's law is invalid, under the assumption that electrodynamic potentials are more fundamental than fields. This leads in general case to $\nabla \cdot \mathbf{E} \neq 0$ even in vacuum for volumes of space outside of electric charges. Therefore longitudinal electric waves are possible in vacuum or free space. Additionally, our recent derivations indicate that Lorenz gauge condition is the cause of vanishing (near-field) longitudinal electric field of dipole radiation commonly found in textbooks. However, these derivations are outside the scope of the present paper, and they will be published in a separate manuscript. In short, application of Gauss's law and/or of Lorenz gauge condition leads to ``disappearance'' of the longitudinal electric waves.

We hope that soon our theoretical results will be checked experimentally and either verified or rejected. However we speculate that Nikola Tesla might have worked already experimentally with these longitudinal electric waves. 

\section{Conclusion}\label{sec:conclusion}

The main contributions of the paper are the following. Firstly, radiation of the infinitesimal electric dipole with oscillating charge magnitudes, but fixed charge positions, was generalized for $\frac{d}{\lambda} \in \mathbb{N}$. Secondly, radiation due to electric current wave traveling in only one direction was obtained. Most of the linear antenna theories consider standing wave currents, which are composed of two current waves traveling in the opposite directions. Thirdly, contribution of the linear charge densities to radiation, generated along the thin wire due to sinusoidal current, was considered. Finally, power generated by longitudinal waves was calculated and compared with the transverse power.

In this work existence of longitudinal electric waves, radiated by the electric dipole when charge separation distance is much larger than wavelength, has been proven theoretically within the framework of classical electrodynamics. By using the assumptions $\lambda \ll d \ll r$, the new results for radiation of an electric dipole were obtained, which are not studied in classical electrodynamics. These results generalize and extend the classical solution, and they indicate that under the above assumptions the electric dipole emits both long-range longitudinal electric and transverse electromagnetic waves. For a specific values of the dipole system parameters the longitudinal and transverse electric fields are displayed. Total power emitted by electric and electromagnetic waves are calculated and compared. It would be very interesting to observe these phenomena in experiments. Longitudinal electric waves could revolutionize wireless communication, remote power transmission and 5G communication technologies.

\section*{Acknowledgments}

We would like to thank our family and our supervisors Prof. Atakan Varol and Prof. Theodoros Tsiftsis for their support.

\section{Appendix I}\label{sec:appendix}

Here we present the derivation of integral expressions. The integral of the second term in  (\ref{eqn:sine}) can be written as

\begin{equation}\label{eqn:appendix1}
\begin{aligned}
& \int_{-\frac{d}{2}}^{\frac{d}{2}} \sin(\frac{2 \pi z}{\lambda}) \cos(\omega(t-\frac{\lrcurs}{c})) \frac{1}{\lrcurs} dz = \\
& \int_{-\frac{d}{2}}^{0} \sin(\frac{2 \pi z}{\lambda}) \cos(\omega(t-\frac{r}{c}(1 + \frac{z}{r}\cos{\theta}))) \frac{1}{r} dz + \\
& \int_{0}^{\frac{d}{2}} \sin(\frac{2 \pi z}{\lambda}) \cos(\omega(t-\frac{r}{c}(1 - \frac{z}{r}\cos{\theta}))) \frac{1}{r} dz  = \\
& \frac{1}{r} \cos(\omega(t-\frac{r}{c}))\Big( \int_{-\frac{d}{2}}^{\frac{d}{2}} \sin(\frac{2\pi z}{\lambda})\cos(\frac{\omega z}{c}\cos{\theta}) dz \Big) + \\
& \frac{1}{r} \sin(\omega(t-\frac{r}{c}))\Big(  \int_{-\frac{d}{2}}^{0} \sin(\frac{2\pi z}{\lambda})\sin(\frac{\omega z}{c}\cos{\theta}) dz - \int_{0}^{\frac{d}{2}} \sin(\frac{2\pi z}{\lambda})\sin(\frac{\omega z}{c}\cos{\theta}) dz \Big)
\end{aligned}
\end{equation}
Each of the three integral terms in the last expression is evaluated below: 
\begin{equation}\label{eqn:integ7}
\begin{aligned}
& \int_{-\frac{d}{2}}^{\frac{d}{2}} \sin(\frac{2\pi z}{\lambda})\cos(\frac{\omega z}{c}\cos{\theta}) dz = 0
\end{aligned}
\end{equation}
\begin{equation}\label{eqn:integ10}
\begin{aligned}
& \int_{-\frac{d}{2}}^{0} \sin(\frac{2\pi z}{\lambda})\sin(\frac{\omega z}{c}\cos{\theta}) dz = \\
& \frac{\lambda}{4\pi}\Big( \frac{1}{1-\cos{\theta}} \sin(\frac{\pi d}{\lambda}(1-\cos{\theta})) - \frac{1}{1+\cos{\theta}} \sin(\frac{\pi d}{\lambda}(1+\cos{\theta})) \Big)
\end{aligned}
\end{equation}
\begin{equation}\label{eqn:integ11}
\begin{aligned}
& \int_{0}^{\frac{d}{2}} \sin(\frac{2\pi z}{\lambda})\sin(\frac{\omega z}{c}\cos{\theta}) dz = \\
& \frac{\lambda}{4\pi}\Big( \frac{1}{1-\cos{\theta}} \sin(\frac{\pi d}{\lambda}(1-\cos{\theta})) - \frac{1}{1+\cos{\theta}} \sin(\frac{\pi d}{\lambda}(1+\cos{\theta})) \Big)
\end{aligned}
\end{equation}
If we substitute the last three equations into (\ref{eqn:appendix1}), then we get 
\begin{equation}\label{eqn:appendix1_result}
\begin{aligned}
& \int_{-\frac{d}{2}}^{\frac{d}{2}} \sin(\frac{2 \pi z}{\lambda}) \cos(\omega(t-\frac{\lrcurs}{c})) \frac{1}{\lrcurs} dz = 0
\end{aligned}
\end{equation}

The integral of the third term in  (\ref{eqn:sine}) can be written as
\begin{equation}\label{eqn:appendix2}
\begin{aligned}
& \int_{-\frac{d}{2}}^{\frac{d}{2}} \cos(\frac{2 \pi z}{\lambda}) \sin(\omega(t-\frac{\lrcurs}{c})) \frac{1}{\lrcurs} dz = \\
& \int_{-\frac{d}{2}}^{0} \cos(\frac{2 \pi z}{\lambda}) \sin(\omega(t-\frac{r}{c}(1 + \frac{z}{r}\cos{\theta}))) \frac{1}{r} dz + \\
& \int_{0}^{\frac{d}{2}} \cos(\frac{2 \pi z}{\lambda}) \sin(\omega(t-\frac{r}{c}(1 - \frac{z}{r}\cos{\theta}))) \frac{1}{r} dz  = \\
& \frac{1}{r} \sin(\omega(t-\frac{r}{c}))\Big( \int_{-\frac{d}{2}}^{\frac{d}{2}} \cos(\frac{2\pi z}{\lambda})\cos(\frac{\omega z}{c}\cos{\theta}) dz \Big) + \\
& \frac{1}{r} \cos(\omega(t-\frac{r}{c}))\Big(- \int_{-\frac{d}{2}}^{0} \cos(\frac{2\pi z}{\lambda})\sin(\frac{\omega z}{c}\cos{\theta}) dz +  \int_{0}^{\frac{d}{2}} \cos(\frac{2\pi z}{\lambda})\sin(\frac{\omega z}{c}\cos{\theta}) dz \Big)
\end{aligned}
\end{equation}
Each of the three integral terms in the last expression were already evaluated in (\ref{eqn:integ1}) - (\ref{eqn:integ5}). If we substitute these results into (\ref{eqn:appendix2}), then we obtain
\begin{equation}\label{eqn:appendix2_result}
\begin{aligned}
& \int_{-\frac{d}{2}}^{\frac{d}{2}} \cos(\frac{2 \pi z}{\lambda}) \sin(\omega(t-\frac{\lrcurs}{c})) \frac{1}{\lrcurs} dz = \\
& \frac{\lambda}{2\pi r} \cos(\omega(t-\frac{r}{c})) \Bigg( -\frac{1}{1-\cos{\theta}} (1-\cos(\frac{\pi d}{\lambda}(1-\cos{\theta}))) + \frac{1}{1+\cos{\theta}} (1-\cos(\frac{\pi d}{\lambda}(1+\cos{\theta})))  \Bigg) + \\
& \frac{\lambda}{2\pi r} \sin(\omega(t-\frac{r}{c})) \Bigg( \frac{1}{1-\cos{\theta}} \sin(\frac{\pi d}{\lambda}(1-\cos{\theta})) + \frac{1}{1+\cos{\theta}} \sin(\frac{\pi d}{\lambda}(1+\cos{\theta})) \Bigg)
\end{aligned}
\end{equation}
The fourth integral term in (\ref{eqn:sine}) can be written as
\begin{equation}\label{eqn:appendix3}
\begin{aligned}
& \int_{-\frac{d}{2}}^{\frac{d}{2}} \sin(\frac{2 \pi z}{\lambda}) \sin(\omega(t-\frac{\lrcurs}{c})) \frac{1}{\lrcurs} dz = \\
& \int_{-\frac{d}{2}}^{0} \sin(\frac{2 \pi z}{\lambda}) \sin(\omega(t-\frac{r}{c}(1 + \frac{z}{r}\cos{\theta}))) \frac{1}{r} dz + \\
& \int_{0}^{\frac{d}{2}} \sin(\frac{2 \pi z}{\lambda}) \sin(\omega(t-\frac{r}{c}(1 - \frac{z}{r}\cos{\theta}))) \frac{1}{r} dz  = \\
& \frac{1}{r} \sin(\omega(t-\frac{r}{c}))\Big( \int_{-\frac{d}{2}}^{\frac{d}{2}} \sin(\frac{2\pi z}{\lambda})\cos(\frac{\omega z}{c}\cos{\theta}) dz \Big) + \\
& \frac{1}{r} \cos(\omega(t-\frac{r}{c}))\Big(  -\int_{-\frac{d}{2}}^{0} \sin(\frac{2\pi z}{\lambda})\sin(\frac{\omega z}{c}\cos{\theta}) dz + \int_{0}^{\frac{d}{2}} \sin(\frac{2\pi z}{\lambda})\sin(\frac{\omega z}{c}\cos{\theta}) dz \Big)
\end{aligned}
\end{equation}
The three integrals in (\ref{eqn:appendix3}) were already evaluated in (\ref{eqn:integ7})-(\ref{eqn:integ11}). Therefore we obtain
\begin{equation}\label{eqn:appendix3_result}
\begin{aligned}
& \int_{-\frac{d}{2}}^{\frac{d}{2}} \sin(\frac{2 \pi z}{\lambda}) \sin(\omega(t-\frac{\lrcurs}{c})) \frac{1}{\lrcurs} dz = 0
\end{aligned}
\end{equation}




\bibliographystyle{spmpsci}      
\bibliography{bibliography5}


\end{document}